\documentclass[a4paper,3p,twocolumn,sort&compress]{elsarticle}

\usepackage[latin1]{inputenc}
\usepackage{graphicx}
\usepackage{diagbox}
\usepackage{color}    
\usepackage{amsmath}
\usepackage{amssymb}
\usepackage[switch]{lineno}
\graphicspath{{../v5/}{.}}

\begin{document}

\title{Absolute hydrogen depth profiling using the resonant $^{1}$H($^{15}$N,$\alpha\gamma$)$^{12}$C nuclear reaction}

\author[tu]{Tobias P. Reinhardt}

\author[hzdr]{Shavkat Akhmadaliev}

\author[hzdr]{Daniel Bemmerer}
\ead{d.bemmerer@hzdr.de}

\author[hzdr,tu]{Klaus St\"ockel}

\author[hzdr,tu]{Louis Wagner}

\address[tu]{Technische Universit\"at Dresden, Institut f\"ur Kern- und Teilchenphysik, 01062 Dresden, Germany}

\address[hzdr]{Helmholtz-Zentrum Dresden-Rossendorf (HZDR), 01328 Dresden, Germany}

\begin{abstract}
Resonant nuclear reactions are a powerful tool for the determination of the amount and profile of hydrogen in thin layers of material. Usually, this tool requires the use of a standard of well-known composition. The present work, by contrast, deals with standard-less hydrogen depth profiling. This approach requires precise nuclear data, e.g. on the widely used $^{1}$H($^{15}$N,$\alpha\gamma$)$^{12}$C reaction, resonant at 6.4\,MeV $^{15}$N beam energy. Here, the strongly anisotropic angular distribution of the emitted $\gamma$-rays from this resonance has been re-measured, resolving a previous discrepancy. Coefficients of (0.38$\pm$0.04) and (0.80$\pm$0.04) have been deduced for the second and fourth order Legendre polynomials, respectively. In addition, the resonance strength has been re-evaluated to (25.0$\pm$1.5)\,eV, 10\% higher than previously reported. A simple working formula for the hydrogen concentration is given for cases with known $\gamma$-ray detection efficiency. Finally, the absolute approach is illustrated using two examples.
\end{abstract}


\maketitle
\nolinenumbers

\section{Introduction}

Recent developments in hydrogen storage and retrieval techniques for energy technology \cite{Fukai05-Book,Zuettel08-Book} and for mobile applications \cite{Ding11-FP} underline the need for simple, precise and generally applicable techniques for the determination of the hydrogen content in a material. 

There are several different nuclear physics based techniques available to determine and profile hydrogen \cite{Khabibullaev89-Book,SchatzWeidinger96-Book,Nastasi15-Book}. This range of techniques exist both because of the low Coloumb barrier generally favoring any light ion induced nuclear reaction with hydrogen, and because the control and determination of the hydrogen content of metals is one of the earliest and most pervasive problems in nuclear technology \cite{MetalHydrides68-Book}. 

Among the various nuclear physical techniques available \cite{Wilde14,Khabibullaev89-Book,SchatzWeidinger96-Book}, NRRA by the $^{1}$H($^{15}$N,$\alpha\gamma$)$^{12}$C resonance at 6.4\,MeV is a preferred choice. This reaction combines high sensitivity (down to hydrogen concentrations of $10^{18}$\,cm$^{-3}$ for bulk materials) with excellent depth resolution down to a few nm near surfaces \cite{Wilde14}. In addition, it enables the study of hydrogen not only near the surface, but up to $4\,\mu\mathrm{m}$ deep in the bulk \cite{Khabibullaev89-Book}.

However, $^{1}$H($^{15}$N,$\alpha\gamma$)$^{12}$C NRRA hitherto requires the comparison of the sample under study with a standard of known hydrogen concentration \cite{Khabibullaev89-Book,SchatzWeidinger96-Book,Wilde14,Nastasi09-Book,Nastasi15-Book}. The hydrogen content was given by the ratio of reaction yields of the sample under study and the standard, respectively, multiplied with the hydrogen content of the sample which was determined by other means. 

The aim of the present work is to go one step further and enable the use of the 6.4\,MeV $^{15}$N resonance as an absolute tool for hydrogen depth profiling. To this end, the discrepant, 60 year old data on the $\gamma$-ray angular distribution \cite{Barnes52-CJP,Kraus53-PR} are addressed with a precise new measurement. In addition, previous information on the resonance strength that depended on  measurements at just one angle \cite{Zijderhand86-NPA,Becker95-ZPA,Marta10-PRC} or on indirect methods \cite{Leavitt83-NPA} are replaced here with data taken in far geometry at three different angles. 

This work is organized as follows. First, the yield formulae for resonant nuclear reaction analyses are reviewed (sec.\,\ref{sec:Theory}). Then, the angular distribution of the emitted $\gamma$-rays from the resonance in the $^{1}$H($^{15}$N,$\alpha\gamma$)$^{12}$C reaction at 6.4\,MeV $^{15}$N beam energy is re-determined experimentally, both with $^{15}$N and with $^{1}$H beam  (sec.\,\ref{sec:AngDist}). Subsequently, the new angular distribution is used to re-evaluate the strength of the resonance (sec.\,\ref{sec:Strength}). The formulae from sec.\,\ref{sec:Theory} and the data from secs.\,\ref{sec:AngDist} and \ref{sec:Strength} are then used to determine a simple working formula for hydrogen depth profiling (sec.\,\ref{sec:Formula}). Finally, the new formula is applied to an example (sec.\,\ref{sec:Examples}), and a summary is given (sec.\,\ref{sec:Summary}).

\section{Formulae for nuclear resonant reaction analyses}
\label{sec:Theory}

\subsection{Derivation of the yield formula}
\label{subsec:Formula}

For energetically narrow resonances, i.e. $\Gamma\ll\Delta E_{\rm CM}$, it can be shown  \cite{Fowler48-RMP} that the maximum resonant yield is given by 
\begin{equation}\label{eq:MaxYieldIliadis}
Y_{\rm max}^\infty = \frac{\lambda_{\rm res}^2}{2} \frac{\omega\gamma}{\left. \frac{dE}{dx}\right|_{\mathrm{eff,CM}}}
\end{equation}
with $\lambda_{\rm res}$ the de Broglie wavelength (in the center of mass system) of the incident particle at the resonance energy, $\omega\gamma$ the so-called resonance strength \cite{Iliadis07-Book}, and $\left. \frac{dE}{dx}\right|_{\mathrm{eff}}$ the energy loss (stopping power) per disintegrable nucleus in the target per area in the units of eV/(at/cm$^{2}$), also in the center-of-mass system.

For a material consisting of additional atoms $i$ in addition to H atoms, the effective stopping power for a $^{15}$N beam is given by
\begin{eqnarray}\label{eq:EffStoppingPowerCM} 
\left. \frac{dE}{dx}\right|_{\mathrm{eff,CM}} & = & \frac{m_{\rm H}}{m_{\rm H}+m_{\rm 15}} \times \frac{1}{n_{\rm H}}  \times \nonumber \\
 & & \left[
\sum\limits_{i} n_{i} \left.\frac{dE}{dx}\right|_{i} + n_{\mathrm{H}} \left. \frac{dE}{dx}\right|_{\mathrm{H}}
\right]
\end{eqnarray}
with $m_{\rm H}$ and $m_{15}$ the atomic masses of hydrogen and $^{15}$N,  $n_{\rm H}$ and $n_{i}$ the atomic concentrations of H and of the additional atoms $i$, and $\left.\frac{dE}{dx}\right|_{i}$ the stopping power of a $^{15}$N ion by atom $i$.

The simple addition of the stopping powers of the various atoms performed in eq.~(\ref{eq:EffStoppingPowerCM}) is appropriate whenever Bragg's rule \cite{Bragg05-PMag} holds. For incident hydrogen and helium ions on organic materials, corrections to Bragg's rule do apply \cite{Thwaites87-NIMB}, and they may be dealt with by the core-and-bond approach \cite{Ziegler88-NIMB,Nastasi09-Book,Ziegler10-NIMB}. The same is true for $^7$Li, $^{12}$C, and $^{16}$O ions in polymer foils \cite{Miksova14-NIMB}. 

An early comprehensive review of stopping in inorganic materials showed no deviations from Bragg's rule within a few percent precision \cite{Ziegler88-NIMB}. More recent research on these cases either confirm Bragg's rule within $\pm$4\% \cite{Zhang03-PRB,Zhang06-NIMB} or show deviations of up to 7\% \cite{Zhang06-NIMB}.

For the present hydrogen depth determination in inorganic materials, it is assumed that Bragg's rule holds. In addition, it is assumed that the SRIM \cite{Ziegler10-NIMB} stopping powers for $^{15}$N ions are correct \cite{Zier12-NIMB}.

\subsection{Hydrogen content}

The sought-after hydrogen concentration $n_{\rm H}$ may be obtained by solving eq.~(\ref{eq:EffStoppingPowerCM}) for $n_{\rm H}$ and inserting eq.~(\ref{eq:MaxYieldIliadis}):
\begin{eqnarray}
n_{\rm H} & = & \frac{\displaystyle\sum\limits_{i} n_{i} \left.\frac{dE}{dx}\right|_{i}}{\displaystyle\frac{m_{\rm H}+m_{\rm 15}}{m_{\rm H}} \frac{\lambda_{\rm res}^2}{2} \frac{\omega\gamma}{Y_{\rm max}^\infty} - \left.\frac{dE}{dx}\right|_{\mathrm{H}}}  
 \label{eq:HydrogenConc}
\end{eqnarray}

The stopping by the hydrogen atoms $\left.\frac{dE}{dx}\right|_{\mathrm{H}}$, included in the denominator of eq.~(\ref{eq:HydrogenConc}), can be neglected for small concentrations $n_{\rm H}$, more precisely for the case 

\begin{equation}\label{eq:LowHydrogenContent}
n_{\rm H} \left.\frac{dE}{dx}\right|_{\mathrm{H}} \ll  \sum\limits_{i} n_{i} \left.\frac{dE}{dx}\right|_{i}
\end{equation}
If this approximation holds, the hydrogen concentration is directly proportional to the observed yield in NRRA, and it can be obtained by simply rescaling with the observed yield for a standard of known hydrogen concentration:

\begin{equation}\label{eq:Proportionality}
n_{\rm H}^{\rm Sample} =  \frac{Y_{\rm max}^{\rm \infty, \, Sample}}{Y_{\rm max}^{\rm \infty, \, Standard}} \, n_{\rm H}^{\rm Standard} 
\end{equation}
This is the formula given in most recent textbooks and reviews \cite{Khabibullaev89-Book,SchatzWeidinger96-Book,Wilde14,Nastasi09-Book,Nastasi15-Book}, which concentrate on low hydrogen concentrations. However, for high hydrogen concentrations, when approximation (\ref{eq:LowHydrogenContent}) is not valid, the simplified formula (\ref{eq:Proportionality}) leads to deviations, e.g. for TiH$_2$ an underestimation of the true hydrogen content by 27\%. 

This problem may be mitigated by using a standard with a hydrogen concentration that is similar to the hydrogen concentration of the sample under study. Even still, keeping the same example of titanium hydride, determining the hydrogen content of a TiH$_{2}$ sample using a TiH$_{1.5}$ standard and the simplified eq.~(\ref{eq:Proportionality}) would lead to 6\% underestimation of the hydrogen amount in the sample. The effect increases for larger deviations between sample and standard and decreases for higher atomic charge numbers of the hydrogen carrier.

\subsection{Conversion from energy scale to depth scale}

In NRRA, the hydrogen content is determined as a function of $^{15}$N beam energy $E$, where higher beam energies correspond to deeper layers in the material. This beam energy scale may be then converted to a depth $x(E)$ by

\begin{equation}
\label{eq:ConversionLength}
x(E) = \frac{(E-E_{\mathrm{res}})}{\displaystyle\sum\limits_{i} n_{i} \left.\frac{dE}{dx}\right|_{i} + n_{\rm H}\left.\frac{dE}{dx}\right|_{\rm H}}
\end{equation}
where, again, the denominator has to be adapted in case of corrections to Bragg's rule.  


In the following sections \ref{sec:AngDist} and \ref{sec:Strength}, the necessary input parameters for hydrogen concentration analysis are derived from new experimental data. These values and tabulated stopping powers are used in sec.\,\ref{sec:Formula} to propose a simple working formula for hydrogen depth profiling, obviating the need for the problematic approximations inherent in eq.~(\ref{eq:Proportionality}). In section \ref{sec:Examples}, a quantitative example is given, including the error analysis.

\section{Measurement of the angular distribution of the emitted $\gamma$-rays}
\label{sec:AngDist}

The 2$^-$ level at $E_{\rm x}$ = 12530\,keV excitation energy in $^{16}$O can be conveniently accessed in two different ways: First, bombarding a hydrogen target with a $^{15}$N beam of $E_{\rm lab}$($^{15}$N) = 6.4\,MeV, and second, using a $^1$H beam of $E_{\rm lab}$($^1$H) = 0.430\,MeV  incident on a $^{15}$N target. The resonance decays predominantly by emission of an $\alpha$-particle to the $E_{\rm x}$ = 4439\,keV first excited state of $^{12}$C ($T_{1/2}$ = 4$\times$10$^{-14}$\,s), which then decays to the ground state by emitting a 4.4\,MeV $\gamma$-ray. The $^{12}$C$^*$(4439) momentum distribution resulting from the $\alpha$-particle recoil leads to an unavoidable Doppler broadening of the observed $\gamma$-ray. 

NRRA using the $E_{\rm lab}$($^{15}$N) = 6.4\,MeV resonance has been introduced in the 1970s \cite{Lanford76-APL}. It gained wide popularity after it was shown that the total energetic width of the resonance is well below 1\,keV, enabling excellent resolution in depth when determining a hydrogen profile \cite{Maurel83-NIM,Zinke85-ZPA,Osipowicz87-NIMB}.

The angular distribution of the emitted $\gamma$ rays by the $^{1}$H($^{15}$N,$\alpha\gamma$)$^{12}$C resonance has been measured in two different experiments in the early 1950s \cite{Barnes52-CJP,Kraus53-PR}. Those experiments used a $^{1}$H beam and solid targets with isotopically enriched $^{15}$N to populate the resonance. Scintillation detectors covering an opening angle of 30$^\circ$ and 36$^\circ$, respectively, were used \cite{Barnes52-CJP,Kraus53-PR}, and the angular corrections by these two works differ by up to 13\%. The only measurement of the angular distribution with a $^{15}$N beam gives no experimental details and reports the results only in graphical form \cite{Horn88-NIMB}.

In light of the differences between Refs.~\cite{Barnes52-CJP,Kraus53-PR}, in the present work, the angular distribution is re-measured, additionally providing both $^{1}$H and $^{15}$N beam data in one and the same setup.

\begin{figure}
\includegraphics[trim=16cm 0 0 0,clip, width=\columnwidth]{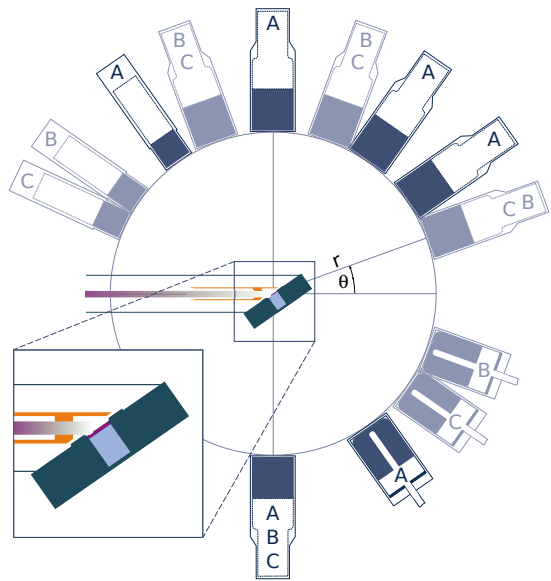}
\caption{\label{fig_setup}(Color online) Schematic cut-away view of the experimental setup.
			    The characters denote the three different configurations (cf. Table \ref{tab_setup}).
			    The inset shows details of  the target chamber, the target (purple), electron suppression (orange), cooling water (light blue), and the beam.}
\end{figure}

\begin{table}
\resizebox{\columnwidth}{!}{%
\begin{tabular}{lcccrcc}
Config. & \multicolumn{5}{c}{LaBr$_{3}$} & HPGe \\ \cline{2-6}
 & \multicolumn{4}{c}{$3''\!\times3''$} & $2''\!\times2''$  &  \\
\hline
A & -90$^\circ$ & 35$^\circ$ & 55$^\circ$ & 90$^\circ$  & 125$^\circ$ & -55$^\circ$ \\
B & -90$^\circ$ & 20$^\circ$ & 70$^\circ$ & 110$^\circ$ & 145$^\circ$ & -20$^\circ$ \\
C & -90$^\circ$ & 20$^\circ$ & 70$^\circ$ & 110$^\circ$ & 155$^\circ$ & -35$^\circ$ \\
\end{tabular}
}%
\caption{\label{tab_setup}
    Angles covered by the $\gamma$ detectors in the three configurations. Positive angles are above the target in Fig.\,\ref{fig_setup}, negative angles below.
   }
\end{table}

\subsection{Beam line and target chamber}
\label{subsec:AngDist_ExpSetup}

The irradiations have been performed at the 3\,MV Tandetron accelerator of Helmholtz-Zentrum Dresden-Rossendorf, Germany \cite{Friedrich96-NIMA}. The $^{1}$H$^{+}$ and $^{15}$N$^{2+}$ beams were provided by a cesium sputter ion source.

The beam from the Tandetron, after being bent by 10$^{\circ}$, successively passes electrostatic quadrupoles and horizontal and vertical deflector units before entering the target chamber (Fig.\,\ref{fig_setup}). There, about 30-50\% of the beam current are absorbed on a water-cooled collimator with an opening of 5\,mm diameter. The beam then passes a 30\,mm long copper tube, negatively biased for secondary electron suppression, that extends to within 2\,mm of the target surface (Fig.\,\ref{fig_setup}). The total charge impinging on the target was measured using an Ortec 439 digital current integrator connected to the electrically insulated target holder.

During the irradiations, the beam line and target chamber were kept under high vacuum, with typical pressures $(2-5)\times10^{-7}$\,mbar.

\begin{figure}
\includegraphics[width=\columnwidth,trim=0 0 14mm 0,clip]{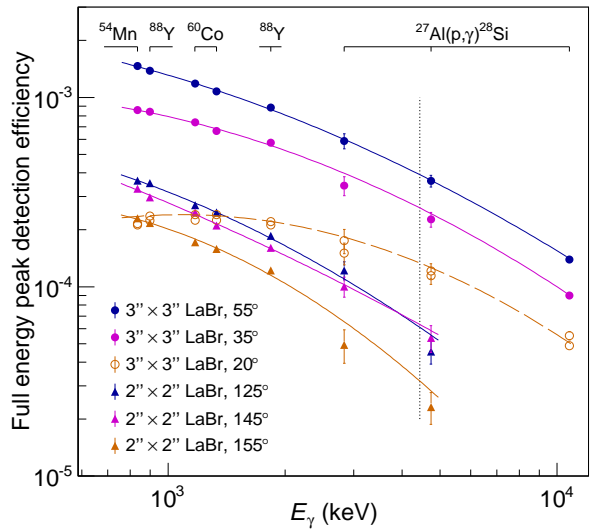}
\caption{\label{fig_efficiency}(Color online)
	$\gamma$-ray full energy peak detection efficiency data. The curves show empirical parametrizations of the efficiency curve for selected LaBr$_{3}$ detectors. The vertical dashed line corresponds to 4.439\,MeV, the $\gamma$-energy of relevance for NRRA by the $^1$H($^{15}$N,$\alpha\gamma$)$^{12}$C reaction.
	}
\end{figure}

\subsection{Targets}
\label{subsec:AngDist_Target}

Two different targets have been used, each based on a tantalum backing of 27\,mm diameter and 0.22\,mm thickness. The targets were mounted tilted by $55^{\circ}$ with respect to the beam axis and directly watercooled during the irradiations. 

For the direct kinematics measurement, a 400\,nm layer of TiN was deposited by means of reactive sputtering on top of the Ta backing using gas of natural isotopic abundance containing 0.4\% of the $^{15}$N isotope.

For the inverse kinematics measurement, a 300\,nm layer of zirconium was evaporated on the Ta backing. Successively, hydrogen was implanted with energies of 15, 10, and 5\,keV and weighted fluences of $3.3$, $1.0$, and $1.3\times10^{17}$\,atoms$/$cm$^{2}$, respectively \cite{Stoeckel15-EPJ}.

\subsection{Detectors}
\label{subsec:AngDist_Detectors}

\begin{figure*}[t]
\includegraphics[width=\textwidth]{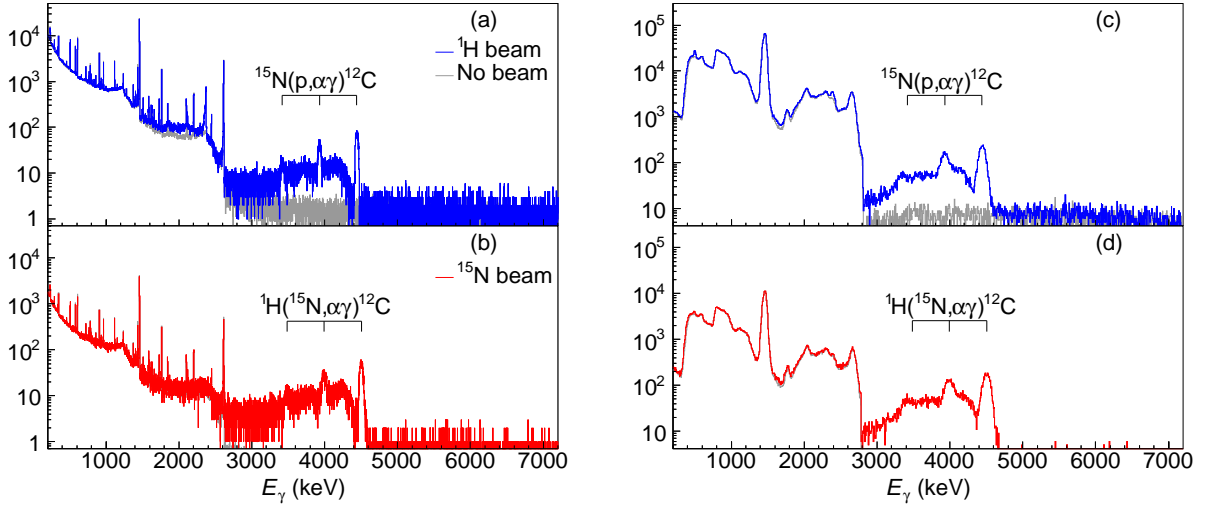}%
\caption{\label{fig_spectra}(Color online)
	$\gamma$-ray spectra on top of the resonance. HPGe detector at -55$^\circ$, $^1$H beam (a) and $^{15}$N beam (b). Arrows denote the signature of the 4439\,keV $\gamma$-ray (full energy, single escape, double escape peaks). The no-beam background is shown in gray, scaled for equal livetime. LaBr$_3$ detector at 55$^\circ$, $^1$H beam (c) and $^{15}$N beam (d).
	}
\end{figure*}

The emitted $\gamma$-rays were observed by six detectors. The detectors were placed in the plane of the beam, mounted atop steel bands connected to an axis below the center of the target \cite{Dietz15-Paris}. The distance between target center and end cap of the detector was 30\,cm. Altogether 12 angles and one reference angle were covered by arranging the detectors in three configurations called A, B, and C, respectively (Fig.\,\ref{fig_setup}, Table~\ref{tab_setup}). 

Four $3''\!\times 3''$ lanthanum bromide (LaBr$_{3}$, cerium doped, trade name Brillance 380) scintillation detectors, one $2''\!\times 2''$ LaBr$_{3}$, and one high-purity germanium (HPGe) detector of 100\,\% relative \cite{Gilmore08-Book} efficiency were used, covering an opening angle of $14^{\circ}$, $10^{\circ}$, and $15^{\circ}$, respectively. The $3''\!\times 3''$ LaBr$_{3}$ scintillators were read out by Hamamatsu R11973 photomultiplier tubes (PMTs), the $2''\!\times 2''$ by a Hamamatsu R2038 PMT.

The first $3''\!\times 3''$ LaBr$_{3}$ was used at a fixed position of -90$^\circ$, as a standard to connect runs in different configurations with each other. The positions of the other detectors are given in Table~\ref{tab_setup}.
All angles are given in the laboratory system, with respect to the direction of travel of the beam.

\subsection{Electronics and data aquisition}
\label{subsec:AngDist_DAQ}

Two different data acquisition (DAQ) systems were used: For the LaBr$_{3}$ detectors, the signals from the PMT anodes were split and the resulting two branches connected to a CAEN V965 charge-to-digital-converter (QDC) and a constant fraction discriminator (CFD), respectively. The trigger and the conversion gate for the QDC were generated in a CAEN V1495 field programmable gate array (FPGA) from the logical OR of the five CFD outputs. The dead time of the LaBr$_{3}$ DAQ system was estimated from the accepted-trigger to trigger ratio 
to be 2\,\%.

Due to the high light yield of the innovative LaBr$_3$ scintillator \cite{Knoll10-Book} and the large efficiency of the PMTs used, saturation effects changed the PMT gain for $\gamma$-ray energies above 3\,MeV. For the graphical representation the pulse height data were thus re-calibrated and re-sampled to give a linear gain. 

For the HPGe detector, the energy signals were amplified and shaped using an Ortec 671 spectroscopic amplifier and digitized and recorded using a histogramming Ortec 919E analog-to-digital converter (ADC) and multichannel buffer unit. A dead time of typically 0.7\% was derived by the Gedcke-Hale method \cite{Jenkins81-Book} for the HPGe DAQ.

\subsection{$\gamma$-ray detection efficiency}
\label{subsec:AngDist_Efficiency}

\begin{figure*}[tb]
\includegraphics[width=\textwidth]{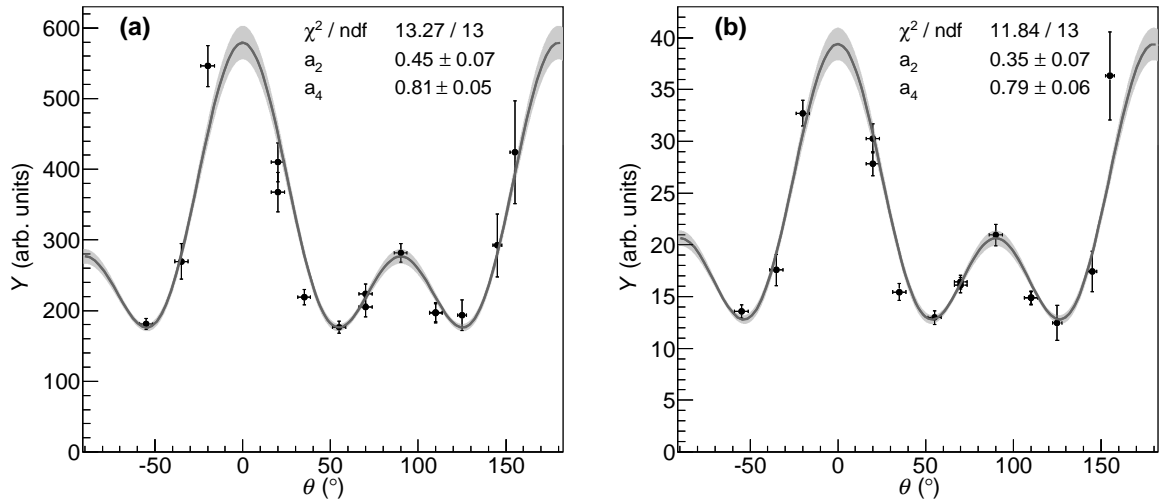}
\caption{\label{fig_ang_dist}
	Angular distribution of the $\gamma$-ray yield and fit to the data for direct (a) and inverse kinematics (b).
	The $1\sigma$ uncertainty band of the fit is shown in gray.
	}
\end{figure*}

For an absolute yield determination, it is important to precisely know the $\gamma$-ray detection efficiency. Therefore, instead of using calculated efficiency values with their inherent uncertainties, here a different approach is adopted  \cite{Marta10-PRC,Schmidt13-PRC,Depalo15-PRC}. Using calibrated radioisotope sources and nuclear reactions, the $\gamma$-ray detection efficiency is obtained up to 10.76\,MeV directly from experimental data. As a first step, the full energy peak detection efficiency was determined using $^{60}$Co, $^{54}$Mn, and $^{88}$Y activity standards with an activity uncertainty of 0.5\%-1.0\%.
In order to extend the curve (Fig.~\ref{fig_efficiency}) to higher energies, in a second step the $E_{\rm p} = 992$\,keV resonance of the $^{27}$Al(p,$\gamma$)$^{28}$Si reaction was used. 
The well-known ratio of emission rates of the 1779- and 10764-keV $\gamma$-rays \cite{Zijderhand90-NIMA}, the known branching ratios \cite{Anttila77-NIM} for the $\gamma$ lines at 2839\,keV and 4743\,keV, and the known angular distributions \cite{Anttila77-NIM} were used for this purpose. 

For some of the $3''\!\times3''$ detectors, the positions were kept the same in configurations B and C, in order to check the reproducibility of the efficiency curve (Fig.~\ref{fig_efficiency}).
Several example efficiency curves are shown in Fig.\,\ref{fig_efficiency}, showing generally similar slopes. The normalization is lower for the $2''\!\times2''$ detector due to its lower active volume. At the lowest angle used here, 20$^\circ$, the curve has a different shape due to a high amount of passive material ($\sim$6\,cm of water, aluminium, and stainless steel) passed by the $\gamma$-rays.  

The resulting 4439\,keV efficiency from the parametric fit of the data at higher and lower $\gamma$-ray energies has 3\% uncertainty for the $3''\!\times3''$ detectors and up to 11\% uncertainty, depending on the configuration, for the $2''\!\times2''$ LaBr$_{3}$.
For the HPGe detector with its better energy resolution, additional $^{27}$Al(p,$\gamma$)$^{28}$Si lines could be used, and the efficiency uncertainty was 2\% for configurations A and B and 8\% for configuration C, where the $^{27}$Al(p,$\gamma$)$^{28}$Si spectrum was accidentally overwritten.

\subsection{Experimental procedure}
\label{subsec:AngDist_Procedure}

As a first step, for a given beam ($^{15}$N or $^1$H), the beam energy was set to the top of the yield curve, $E_{\rm lab}(^{15}{\rm N})$ = 6.6\,MeV and $E_{\rm lab}(^{1}{\rm H})$ = 0.462\,MeV, respectively. Then, irradiations were performed subsequently in the three configurations A, B, and C. The beam current on target was 0.7-1.8\,$\mu$A for $^{15}$N$^{2+}$ (corrected for secondary electron effects) and 4-6 \,$\mu$A for $^{1}$H$^+$. 

The yields from the three configurations are connected to each other by normalizing to the yield of the detector at -90$^\circ$ that was kept at the same position throughout the experiment. Possible degradations in the target under the intense ion beam bombardment therefore affect the yield of the reference detector and the yield at the angle under study by exactly the same factor. Indeed it was found that the hydrogen implanted Zr target showed a 23\% decrease of the yield at -90$^\circ$ over the course of the experiment. Because of the normalization to the -90$^\circ$ detector, this decrease does not affect the results. The typical irradiation time for the angular distribution measurement was several hours per data point.
The TiN target, instead, remained stable under bombardment. 

\subsection{Interpretation of the in-beam $\gamma$-ray spectra}
\label{subsec:AngDist_GammaSpectra}

The observed $\gamma$-ray spectra in the HPGe detector show no significant beam-induced lines except for the ones from the reaction under study (Fig.\,\ref{fig_spectra}). The observed $\gamma$-ray energy at the 55$^\circ$ forward angle subtended by the HPGe in configuration B is higher in the $^{15}$N beam case than for $^1$H beam. This is due to the higher Doppler shift because of the higher velocity of the center of mass for $^{15}$N beam. For the same reason, the Doppler broadening of the peak is larger.

The LaBr$_3$ detectors show the same general picture as the HPGe (Fig.\,\ref{fig_spectra}). However, the LaBr$_3$ energy resolution is lower than for HPGe, as expected. 

For the analysis of the angular distribution, only the full energy peaks were used. The laboratory background was subtracted, scaled for equal livetime. 
The statistical uncertainty of the resultant $\gamma$-yield was 1-5\% for the $3''\!\times3''$ LaBr$_{3}$, 2-12\% for the smaller $2''\!\times2''$ detector, and up to 3\% for the HPGe.

\subsection{Resulting angular distribution}

The 4439\,keV yields, after normalization by the $-90^{\circ}$ detector to one common scale given by configuration A, were plotted together (Fig.\,\ref{fig_ang_dist}). The y axis error bars in Fig.\,\ref{fig_ang_dist} reflect the statistical uncertainty, the efficiency error, and the statistical uncertainty of the -90$^\circ$ normalization. The x axis error bars correspond to $1/\sqrt{12}$ of the full angle subtended by the detector crystals, i.e. for the purpose of the fit it is assumed that the full angle corresponds to 92\% coverage for a normal distribution.

The y axis normalization of Fig.\,\ref{fig_ang_dist} has first been determined by fitting each of the two yield curves with the parameterization

\begin{equation}\label{eq:AngDist}
W(\theta) = a_0 \left[ 1 + a_2 P_2(\cos\theta) + a_4 P_4(\cos\theta) \right]
\end{equation}
where $P_{2,4}$ are the second and fourth order Legendre polynomials and $a_{0,2,4}$ are parameters to be determined. Then, the arbitrary normalization is included in parameter $a_0$.

In a second step, for each of the two kinematics $a_0$ was fixed at its previously fitted value, and the fit was repeated varying $a_{2,4}$. The resulting $a_{2,4}$ parameters and their errors are shown in the insets in Fig.\,\ref{fig_ang_dist}. The $a_{2,4}$ results are mutually consistent for both $^{1}$H and $^{15}$N beams, and their weighted averages are then adopted (Table~\ref{tab_parameters}).

\begin{table}[tb]
\caption{\label{tab_parameters}
	Angular distribution coefficients $a_{2,4}$ for eq.~(\ref{eq:AngDist}). In addition, the ratio of angular corrections $W(\theta)$ is given for 90$^\circ$ and 55$^\circ$.
	}
\resizebox{\columnwidth}{!}{%
\begin{tabular}{lllll} 
 & \multicolumn{1}{c}{$a_{2}$} & \multicolumn{1}{c}{$a_{4}$} & $\displaystyle\frac{W(90^\circ)}{W(55^\circ)}$  \\
\hline
Barnes \textit{et al.} \cite{Barnes52-CJP} & $0.24\pm0.11$ & $0.96\pm0.06$ & 1.98 \\ 
Kraus \textit{et al.} \cite{Kraus53-PR} & 0.33  &  0.80 & 1.65  \\
Present work & $0.38\pm0.04$  & $0.80\pm0.04$ & 1.60$\pm$0.06 \\ \hline \hline
\end{tabular}
}
\end{table}

In order to facilitate the comparison with the present data, the angular distribution coefficients given in the literature \cite{Barnes52-CJP,Kraus53-PR} have been converted to the nowadays adopted Legendre parameterization (Table~\ref{tab_parameters}). The only inverse-kinematics work \cite{Horn88-NIMB} did not provide the data or parameterization in numerical form but reported its data to be consistent with Kraus {\it et al. } \cite{Kraus53-PR}. 

For one pair of angles $\theta$, the ratio of the angular corrections $W(\theta)$ have also been computed  (Table~\ref{tab_parameters}), in order to give a tangible impression of the anisotropy and facilitate the comparison. The present result is less anisotropic than what was found by Barnes {\it et al.} \cite{Barnes52-CJP} but consistent with the Kraus {\it et al.} result \cite{Kraus53-PR}. Due to the smaller opening angles of the detectors used here, the present result is less prone to systematic errors than these previous works \cite{Barnes52-CJP,Kraus53-PR} and should be recommended for further use.

\section{Re-determination of the resonance strength}
\label{sec:Strength}

For the purposes of hydrogen depth profiling compared to a standard, it is sufficient to know that the cross section on top of the resonance is much larger than the off-resonant cross section, i.e. the ratio of resonant to off-resonant cross section. This assumption has previously been proven experimentally \cite{Horn88-NIMB}. Recently, new data and extrapolations became available for the off-resonant cross section \cite{Wilde05-NIMB,Imbriani12-PRC,Imbriani12-PRC-Erratum}. 

For the purposes of standardless hydrogen concentration, however, the resonant cross section $\sigma_{\rm res}$ must be known. Here, instead of $\sigma_{\rm res}$, the resonance strength $\omega\gamma$ is used. 

The total width $\Gamma$ of the resonance had originally been assumed to be as high as 900\,eV \cite{Schardt52-PR,Hebbard60-NP}. Later it was discovered that it was much lower, of the order of 120\,eV, greatly spurring interest in its use for applied physics \cite{Maurel83-NIM}. The most precise width value to date, $\Gamma$ = $(124\pm17)$\,eV, has been obtained using a proton beam on a frozen nitrogen target \cite{Osipowicz87-NIMB}.

For the strength, a number of previous measurements exist (Table~\ref{table:omegagamma}). For the pre-1985 works, the experimental results given are converted to a resonance strength $\omega\gamma$ here, in order to facilitate a comparison. The quoted thick-target yield in the early work by Schardt {\it et al.} \cite{Schardt52-PR} has been converted here to a resonance strength using textbook \cite{Iliadis07-Book} formulae. Hebbard \cite{Hebbard60-NP} gives the proton and $\alpha$ widths, this is converted to $\omega\gamma$ here. The integrated $\gamma$-ray yield given by Gorodetzky {\it et al.} \cite{Gorodetzky68-NPA} has been converted to a strength. Leavitt {\it et al.} \cite{Leavitt83-NPA} studied several levels of interest and extracted the proton and $\alpha$-particle widths from an R-matrix fit; from these values $\omega\gamma$ has been obtained. 

Zijderhand and van der Leun \cite{Zijderhand86-NPA} have determined the resonance strength relative to the  $^{15}$N(p,$\alpha\gamma$)$^{12}$C resonance at $E_{\rm lab}(^1{\rm H})$ = 897 keV, using two Ge(Li) detectors in close geometry at 55$^\circ$ angle. Becker {\it et al.} used a gas target, $\alpha$-particle yield measurements at one angle, and a previous $\alpha$-particle angular distribution \cite{Leavitt83-NPA} to determine the strength \cite{Becker95-ZPA}. Marta {\it et al.} \cite{Marta10-PRC} measured the strength relative to the recently developed standard value \cite{Adelberger11-RMP} of $\omega\gamma_{278}$ = (13.1$\pm$0.6)$\times10^{-3}$\,eV for the $E_{\rm p}$ = 278\,keV resonance in the $^{14}$N(p,$\gamma$)$^{15}$O reaction.

\begin{table}[tb]
\resizebox{\columnwidth}{!}{%
\begin{tabular}{|r@{$\pm$}l|l|l|p{5.6cm}|}
\hline
\multicolumn{2}{|c|}{$\omega\gamma$ [eV]} & Ref. & Year & Method \\ \hline
\multicolumn{2}{|c|}{26} & \cite{Schardt52-PR} & 1952 & $\gamma$-ray yield measurement. \\
\multicolumn{2}{|c|}{24} & \cite{Hebbard60-NP} & 1960 & $\gamma$-ray yield measurement. \\
27 & 3 & \cite{Gorodetzky68-NPA} & 1968 & $\gamma$-ray yield measurement. \\
21 & 4 & \cite{Leavitt83-NPA} & 1983 & Based on R-matrix fit. \\
17 & 2 & \cite{Zijderhand86-NPA} & 1986 & $\gamma$-ray yield, relative to resonance at $E_{\rm p}$ = 897 keV, one angle. \\
21.1 & 1.4 & \cite{Becker95-ZPA} & 1995 & $\alpha$-yield measurement, gas target, inverse kinematics, one angle. \\
22.7 & 1.4 & \cite{Marta10-PRC} & 2010 & $\gamma$-yield, relative to $^{14}$N(p,$\gamma$)$^{15}$O resonance at $E_{\rm p}$ = 278 keV, one angle. \\ \hline
25.0 & 1.5 & \multicolumn{2}{c|}{Present} & Re-analysis of data from Ref.~\cite{Marta10-PRC}, using the present angular distribution and three angles instead of one. \\ \hline
\end{tabular}
}
\caption{Total cross section of the $E_{\rm p}$ = 430\,keV resonance in the $^{15}$N(p,$\alpha\gamma$)$^{12}$C reaction from the literature \cite{Schardt52-PR,Hebbard60-NP,Gorodetzky68-NPA,Leavitt83-NPA,Zijderhand86-NPA,Becker95-ZPA,Marta10-PRC} and from the present work, expressed as resonance strength. See text for details.}
\label{table:omegagamma}
\end{table}

Marta {\it et al.} obtained their strength based on the high-statistics yield of one detector placed in close geometry at an angle of 55$^\circ$ with respect to the beam direction \cite{Marta10-PRC}. However, the yield for this close-geometry detector does not agree within one standard deviation with the predicted value using the present, newly determined angular distribution. Summarizing the four post-1980 works \cite{Leavitt83-NPA,Zijderhand86-NPA,Becker95-ZPA,Marta10-PRC}, one uses an R-matrix fit for the strength \cite{Leavitt83-NPA}, one measures the (strongly anisotropic) $\alpha$-yield at just one angle, and the final two measure the (strongly anisotropic) $\gamma$-yield at just one angle and in close geometry.

In order to improve this unsatisfactory literature situation, the resonance strength is re-determined here from the original Marta {\it et al.} data, using three detectors located at $\pm$127$^\circ$ and 90$^\circ$ in far geometry instead of the previous one, close-geometry detector. By using far-geometry detectors, the solid angle subtended by each detector is reduced, improving the systematic uncertainty by sacrificing some statistics. The resulting strength from this present re-analysis of the Marta {\it et al.} data is 10\% higher, $\omega\gamma = (25.0\pm1.5)$\,eV (Table~\ref{table:omegagamma}). 

\section{Working formula for hydrogen analyses}
\label{sec:Formula}

\begin{table*}[bth]
\caption{\label{tab_detection_systems}
	Product $\eta\times W'(\theta)$ of  detection efficiency and angular correction from a GEANT4 simulation for several different detectors, angles $\theta$ to the beam direction, and distances $d$ to the target.
	}
\resizebox{\textwidth}{!}{%
\begin{tabular}{r*{9}{p{0.09\columnwidth}}}
\diagbox{$\theta$}{Detector} & \multicolumn{3}{|c|}{$3''\!\times3''$ BGO} & \multicolumn{3}{c|}{$4''\!\times4''$ BGO} & \multicolumn{1}{p{0.25\columnwidth}|}{$10''\!\times10''$ NaI} & \multicolumn{1}{p{0.2\columnwidth}|}{$4\pi$ NaI} & \multicolumn{1}{p{0.2\columnwidth}|}{$4\pi$ BGO} \\
\hline
d (mm) & 10 & 20 & 30 & 10 & 20 & 30 & 70 & & \\ 
\hline
$0^{\circ}$ & 0.1196 & 0.0877 & 0.0664 & 0.1734 & 0.1361 & 0.1091 & 0.1075 & 0.4507 & 0.3218\\ 
$45^{\circ}$ & & & & & & 0.0725 & 0.0727 & & \\
$55^{\circ}$ & & & & & & 0.0653 & 0.0657 & & \\ 
$90^{\circ}$ & & & & & & 0.0592 & 0.0588 & & \\ 
$135^{\circ}$ & & & & & & 0.0726 & 0.0728 & & \\ 
\end{tabular}
}
\end{table*}
\begin{table}[b]
\caption{\label{tab_stopping}
Stopping power $\left.\frac{dE}{dx}\right|_{i}$ for $6.4$\,MeV $^{15}$N$^{+,2+}$ ions in different materials $i$ from the SRIM \cite{Ziegler10-NIMB} software, in units of $\rm eV / (10^{15} at/cm^2)$.
	}
\begin{tabular}{lr}
Target $i$ & Stopping power \\ 
\hline
H (solid) & 61.1 \\ 
Ti & 445 \\
TiN & 445+195 = 640 \\
Ta & 632 \\ 
Si & 279 \\ 
GaN & 422+195 = 617 \\ 
\end{tabular}
\end{table}

In the present section, a simple working formula for the hydrogen content in frequently investigated materials in a given NRRA setup is developed, using the full eq.~(\ref{eq:HydrogenConc}).

The yield $Y_{\rm max}^\infty$ at the plateau of the resonance can be obtained by the experimental observable, the number of observed counts per unit charge (in Coulombs) on target $Y_{\rm Q}$, by

\begin{equation}\label{eq:YieldInfinity}
Y_{\rm max}^\infty = \frac{Y_{\rm Q} \times qe}{\eta \times W'(\theta)}
\end{equation}
where $qe$ is the electrical charge of one beam particle (typically one or two elementary charges, depending on whether $^{15}$N$^+$ or $^{15}$N$^{2+}$ beam is used), and $\eta$ the efficiency of the detector used for detecting the 4.4\,MeV $\gamma$ rays. $W'(\theta)$ is a modified form of the angular correction from eq.~(\ref{eq:AngDist}), without the normalization and corrected for the attenuation of the anisotropy by the fact that the detector is not infinitely small and thus averages over some angular range:

\begin{equation}\label{eq:AngDistRose}
W'(\theta) = 1 + a_2 Q_2 P_2(\cos\theta) + a_4 Q_4 P_4(\cos\theta)
\end{equation}
In principle, the attenuation coeffients $Q_{2,4}\in[0;1]$ may be analytically calculated for simple geometries \cite{Rose53-PR}. For the present purposes, they are instead determined from a Monte Carlo simulation in the GEANT4 \cite{Agostinelli03-NIMA} framework that directly produces the product $\eta\times W'(\theta)$, for the detection of the full energy peak of the 4.439\,MeV $\gamma$ ray (Table~\ref{tab_detection_systems}). 

For detectors and geometries not included in Table~\ref{tab_detection_systems}, $\eta\times W'(\theta)$ may be determined in two different ways. First, developing a Monte Carlo simulation describing the detector geometry, and including the un-attenuated angular distribution $W(\theta)$ from eq.~(\ref{eq:AngDist}). Second, by measuring the $\gamma$-ray detection efficiency at 4.439\,MeV by using radioactive sources and nuclear reactions (sec.~\ref{subsec:AngDist_Efficiency}) and then using the attenuated angular distribution $W'(\theta)$ from eq.~(\ref{eq:AngDistRose}) using calculated \cite{Rose53-PR} attenuation coeffients.

The simulated values of $\eta\times W'(\theta)$ in Table~\ref{tab_detection_systems} have been motivated by a selection of setups at various laboratories worldwide that are used for NRRA hydrogen depth profiling with a $^{15}$N beam. Only setups with a sufficiently detailed description of the geometry were included \cite{Amsel98-NIMB,Wilde14,Liedke15-JAP,Uhrmacher05-JAC,Traeger11,Torri94}. Passive materials between target and detector have been neglected:
\begin{itemize}
\item In Paris \cite{Amsel98-NIMB}, an unshielded $4''\!\times4''$ BGO placed at $0^{\circ}$ in 20\,mm distance was used for NRRA.
\item A similar setup with the same detector geometry but 30\,mm distance to the sample is located at the 5 MV tandem in Tokyo \cite{Wilde14}.
\item In Dresden, the AIDA2 upgrade of the AIDA setup \cite{Liedke15-JAP} includes a $3''\!\times3''$ BGO crystal at $0^{\circ}$ for NRRA and a 5-axes manipulator serving as sample holder enabling to use different distances to the detector.
\item At the 3 MV pelletron in G\"ottingen \cite{Uhrmacher05-JAC}, a low-level $\gamma$ counting setup uses a $10''\!\times10''$ NaI(Tl) detector with anti-muon veto at $90^{\circ}$ with 70\,mm distance\cite{Damjantschitsch83} to the target.
\item In Bochum \cite{Traeger11}, a $12''\!\times12''$ $4\pi$ NaI(Tl) detector with a bore hole of 35\,mm diameter and the sample in the very center \cite{Spyrou07} is used for NRRA.
\item Another bore hole detector is used in Helsinki \cite{Torri94}, comprising an annular BGO cystal of 200\,mm length and 35\,mm radial thickness surrounding an opening of 89\,mm diameter.
\end{itemize}

In order to facilitate the discussion, an abbreviation for the constant factors included in eq.~(\ref{eq:HydrogenConc}) is introduced, also including the elementary charge from eq.~(\ref{eq:YieldInfinity}):
\begin{eqnarray}
\varepsilon &=& \frac{1}{e}\left(\frac{m_{\rm H}+m_{15}}{m_{\rm H}} \frac{\lambda_{\rm res}^2}{2} \omega\gamma \right) \nonumber \\
 &=& 2.69 \times 10^{13} \, \rm C^{-1} \, eV / (10^{15} at/cm^2) \label{eq:K}
\end{eqnarray}
Equation~(\ref{eq:HydrogenConc}) can then be expressed as:

\begin{equation}\label{eq:HydrogenConc2}
n_{\rm H} = \frac{\displaystyle\sum\limits_{i} n_{i} \left.\frac{dE}{dx}\right|_{i}}{\displaystyle \frac{\varepsilon
\times \eta \times W'(\theta)}{q \times Y_{\rm Q}} - \left.\frac{dE}{dx}\right|_{\mathrm{H}}}  
\end{equation}
The stopping power $\left.\frac{dE}{dx}\right|_{i}$ has been tabulated here for several materials with relevance to hydrogen depth profiling (Table~\ref{tab_stopping}), using data from the SRIM \cite{Ziegler10-NIMB} software. The uncertainty for the SRIM stopping power values of ions heavier than beryllium, thus also for $^{15}$N, has been estimated \cite{Ziegler10-NIMB} to be 5.6\%.

The inputs to eq.~(\ref{eq:HydrogenConc2}) are thus:
\begin{itemize} 
\item The concentrations $n_x$ in atoms/cm$^3$ of the various elements in the compound.
\item Stopping powers $\left.\frac{dE}{dx}\right|_{i}$ from Table~\ref{tab_stopping} or SRIM \cite{Ziegler10-NIMB}.
\item $\varepsilon  = 2.69 \times 10^{13} \, \rm C^{-1} \, eV / (10^{15} at/cm^2)$, eq.~(\ref{eq:K}).
\item Angle-weighted efficiency $\eta\times W(\theta)$ (Table~\ref{tab_detection_systems}).
\item Charge state $q$ of the beam ($q$=2 for $^{15}$N$^{2+}$ beam).
\item Experimental yield $Y_{\rm Q}$ in counts per Coulomb incident beam charge.
\item Hydrogen stopping power $\left.\frac{dE}{dx}\right|_{i}$ (Table~\ref{tab_stopping}).
\end{itemize} 
Using these values and eq.~(\ref{eq:HydrogenConc2}), the hydrogen content of a sample under study can be directly determined, without the need for a standard or an approximation requiring low hydrogen content.

\section{Example hydrogen depth profiles}
\label{sec:Examples}

\begin{figure*}
\includegraphics[width=\textwidth]{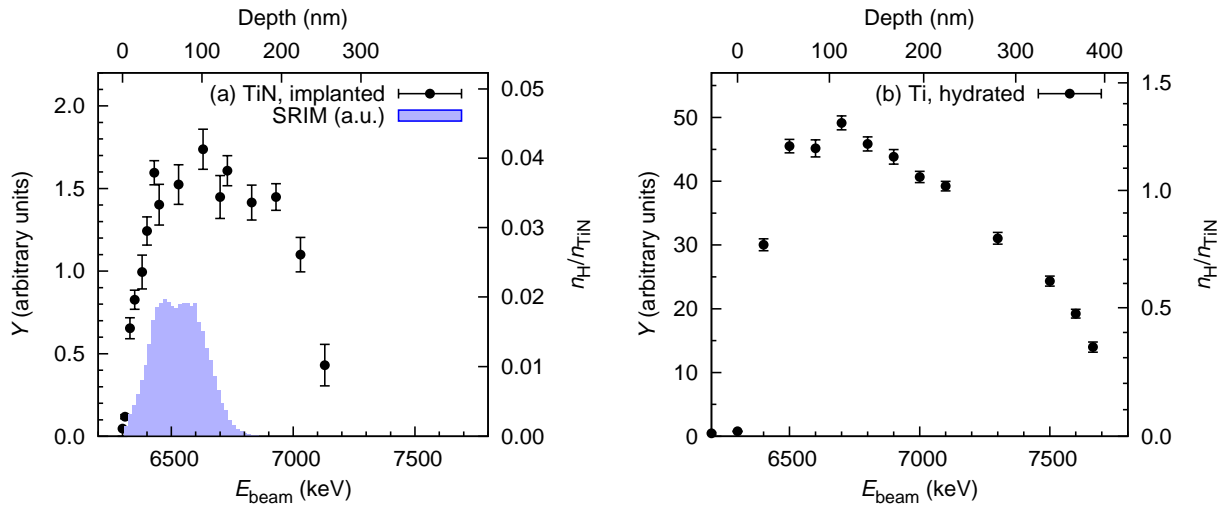}
\caption{\label{fig_target_profiles}
	(Color online) Hydrogen depth profiles of (a) hydrogen implanted in TiN and (b) hydrated Ti. In panel (a), the predicted depth profile from a SRIM simulation using the implantation profile is also shown, scaled arbitrarily.
	 The slightly non-linear behavior of the concentration scale (right y-axis) in panel (b) is caused by the stopping power of hydrogen.
	See text for further details.
	}
\end{figure*}

As an illustration, hydrogen concentration data from an ongoing cross section measurement of the astrophysically relevant $^{12}$C(p,$\gamma$)$^{13}$N reaction \cite{Stoeckel15-EPJ} are shown and analyzed in the following. In the experiment, a solid hydrogen target is irradiated with a $^{12}$C beam, and the $\gamma$ rays are detected with a 60\% relative \cite{Gilmore08-Book} efficiency HPGe detector placed at $55^{\circ}$ angle with respect to the beam, at a distance to the target of $d$ = 34\,mm. 

The challenge in that experiment is to determine the hydrogen content of the target in situ by changing the beam from $^{12}$C to $^{15}$N. To this end, each $^{12}$C irradiation is bracketed by $^{15}$N irradiations before and afterwards, allowing to judge both the relative target stability and the absolute hydrogen content quantitatively. Several different target production schemes were tried, in order to determine which target best sustains the long, high-intensity $^{12}$C irradiations. 
The parameters for eq.~(\ref{eq:HydrogenConc2}) are $q$=2 ($^{15}$N$^{2+}$ beam), $\eta\times W'(\theta)$ = 0.00201$\pm$0.00007. 

The first example used for the present study is a titanium nitride layer of 400\,nm thickness that is deposited on a 0.22\,mm thick tantalum carrier by the reactive sputtering technique. The TiN layer was implanted with 11 and 5\,keV hydrogen ions with weighted fluences of $6.0$ and $3.4\times10^{17}$\,atoms$/$cm$^{2}$, respectively. The sample was cooled by liquid nitrogen during the implantation.

After significant $^{12}$C bombardment, the hydrogen profile of the sample is determined, using eq.~(\ref{eq:ConversionLength}) for the x-axis and eq.~(\ref{eq:HydrogenConc2}) for the y-axis shows a steep rising slope at the target surface and an effective width of the hydrogen profile of about 200\,nm (Fig.~\ref{fig_target_profiles}, left panel). The comparison with the predicted profile from the implantation profile (using SRIM \cite{Ziegler10-NIMB}) shows that the implanted hydrogen was rather mobile and diffused up to a depth of 200\,nm, presumably limited by the effective target temperature at this depth, in equilibrium between heating by the ion beam and the liquid nitrogen cooling of the backing. 

\begin{table*}[t!!]
\caption{\label{tab_target_uncertainty_budget}
	Error budget for the hydrogen concentration.
	}
\begin{center}
\begin{tabular}{lrrr}
Effect & \multicolumn{1}{c}{Relative} & \multicolumn{2}{c}{Contribution} \\
 & error & \multicolumn{2}{c}{to $\Delta n_{\rm H}/n_{\rm H}$} (\%)\\
 & & (a) TiN & (b) Ti \\
\hline
Resonance strength $\omega\gamma$ & 6.0 & 6.1 & 7.2 \\
$\eta \times W'(\theta = 55^{\circ})$ & 3.7 & 3.8 & 4.4 \\
$^{15}$N stopping in Ti and N, $\left.\frac{dE}{dx}\right|_{i}$ & 5.6 & 5.6 & 5.6 \\
$^{15}$N stopping in solid H, $\left.\frac{dE}{dx}\right|_{\mathrm{H}}$ & 5.6 & 0.1 & 1.1 \\
& \\
Total systematic & & 8.4 & 10.2 \\
Statistical (on the plateau) & & 4.6 & 2.1 \\
\end{tabular}
\end{center}
\end{table*}

The second example discussed here consists of a 300\,nm thick layer of titanium, deposited by evaporation on a 0.22\,mm thick tantalum backing. This layer was then hydrated by the following process: It was exposed to a hydrogen atmosphere (normal pressure, constant flow of 10\,liters/hour) and slowly (1\,K/min) heated up to 350$^{\circ}$\,C and held for two hours at the nominal temperature, then slowly cooled to room temperature, all the while maintaining the hydrogen flow. The sample was subsequently used for a two day long irradiation with $^{12}$C beam for an astrophysically motivated study \cite{Stoeckel15-EPJ}. 

The hydrogen profile was determined also for this sample (fig.~\ref{fig_target_profiles}, right panel), again after $^{12}$C bombardment. The absolute hydrogen level for the sample reaches a maximum stoichiometry of TiH$_{1.3}$, similar to the frequently used level of TiH$_{1.5}$ \cite{Stoeckel15-EPJ}. Hydrogen is seen in also beyond the Ti/Ta boundary at 300\,nm, probably due to the effect of the heavy $^{12}$C bombardment mobilizing some hydrogen atoms. Note that the hydrogen concentration for the hydrated titanium sample would be underestimated by 11\% if the stopping power of the hydrogen in eq.~(\ref{eq:HydrogenConc2}) were neglected. 

The error budget is then evaluated. For case (a) shown here, relation~(\ref{eq:LowHydrogenContent}) applies. For case (b) instead, the stopping by hydrogen is not negligible anymore. This latter effect leads to an over-proportional influence of the parameters $\omega\gamma$ and $\eta \times W'(\theta = 55^{\circ})$ entering $\varepsilon$ (Table~\ref{tab_target_uncertainty_budget}, last column). 

The resulting relative uncertainty is dominated in both examples given by the three components resonance strength (sec.\,\ref{sec:Strength}), angular distribution   (sec.\,\ref{sec:AngDist}) and efficiency, and $^{15}$N energy loss in the carrier matrix \cite{Ziegler10-NIMB}. These effects lead to 8\% systematic uncertainty for a sample of low hydrogen content, where approximation (\ref{eq:LowHydrogenContent}) applies, and 10\% systematic uncertainty for a sample with high hydrogen content, where (\ref{eq:LowHydrogenContent}) does not hold. 
In both cases shown here, the statistical uncertainty is negligible when compared to the systematic uncertainty, for a typical running time of 2-3 minutes per data point on the resonance plateau for both cases shown here. 

\section{Discussion and summary}
\label{sec:Summary}

Hydrogen depth profiling by way of nuclear resonant reaction analysis using the $^{1}$H($^{15}$N,$\alpha\gamma$)$^{12}$C resonance at $E_{\rm lab}(^{15}{\rm N})$ = 6.4\,MeV has been re-examined. It was shown that for samples with high hydrogen content, the textbook approach of determining the hydrogen content by scaling with the yield on a standard of known composition \cite{Khabibullaev89-Book,SchatzWeidinger96-Book,Wilde14,Nastasi09-Book,Nastasi15-Book} may lead to deviations, if the  stopping power contribution of the hydrogen atoms is neglected. An absolute determination of the hydrogen content is one way to correct this problem.

Subsequently, two experimental building blocks for such an absolute determination have been provided in the present work: The angular distribution of the emitted $\gamma$-rays from the $^{1}$H($^{15}$N,$\alpha\gamma$)$^{12}$C resonance has been measured, first with $^1$H beam incident on a $^{15}$N target, then with $^{15}$N beam incident on a $^1$H target (used for NRRA hydrogen depth profiling), and Legendre coefficients of $a_2$ = 0.38$\pm$0.04 and $a_4$ = 0.80$\pm$0.04 have been derived from the data. Subsequently, the absolute resonance strength has been re-determined to be (25.0$\pm$1.5)\,eV. 

Finally, a simple working formula for absolute hydrogen depth profile determination has been developed based on the new data and applied to two examples. The present approach and formula are limited to cases where either Bragg's stopping power addition rule holds, or where deviations can be quantified. For cases with unknown deviations to Bragg's stopping power rule, an additional uncertainty has to be taken into account. 

Two examples show that this approach can yield hydrogen profiles with 8-10\% systematic uncertainty, depending on the actual hydrogen concentration. This absolute error bar is higher than the 2\% reproducibility which has been reported previously in an intercomparison exercise involving a low concentration sample \cite{Boudreault04-NIMB}. It is mainly due to the remaining uncertainty in the resonance strength and in the stopping power. Even still, it is hoped that these data and considerations prove to be useful for cases of high hydrogen concentrations and wherever the use of standards is difficult. 

\section*{Acknowledgments}
The authors are indebted to Dr. Arnd Junghans (HZDR) for the loan of the LaBr$_3$ detectors, to Dr. Andreas Wagner, Dr. Roland Beyer, Dr. Konrad Schmidt, and the late Mathias Kempe (HZDR) for support with the data acquisition, to Dr. Oliver Busse (TU Dresden) for the hydration of the titanium sample, to Mario Steinert and Franziska Nierobisch (HZDR) for the production and implantation, respectively, of the H-implanted TiN sample, to Andreas Hartmann (HZDR) for technical support, and to Dr. Jan-Martin Wagner (University of Kiel) for a critical reading of the manuscript. 
--- This work was supported in part by NupNET NEDENSAA (BMBF 05P120DNUG), by GSI F\&E (DR-ZUBE), and by the Helmholtz Association's Nuclear Astrophysics Virtual Institute (HGF NAVI VH-VI-417). L.W. is an associate member of the Helmholtz Graduate School for Hadron and Ion Research HGS-HIRe.

\section*{References}

\providecommand{\noopsort}[1]{}\providecommand{\singleletter}[1]{#1}%

\end{document}